\documentclass[prl,aps,superscriptaddress,twocolumn,showpacs,preprintnumbers,amsmath,amssymb,reprint]{revtex4-1}

\usepackage{graphicx}
\usepackage{amsmath}
\usepackage{amssymb}
\usepackage{color}
\usepackage{dcolumn}
\usepackage{epsfig}
\usepackage{bm}
\usepackage{subfigure}
\usepackage[urlcolor=blue]{hyperref}
\hypersetup{backref, colorlinks=true, linkcolor=blue, citecolor=blue}
\begin{document}

\title{Synthesis and Superconductivity in Yttrium-Cerium Hydrides at Moderate Pressures}

\author{Liu-Cheng Chen}
\affiliation{School of Science, Harbin Institute of Technology, Shenzhen 518055, China}
\affiliation{Center for High Pressure Science and Technology Advanced Research, Shanghai 201203, China}

\author{Tao Luo}
\affiliation{School of Science, Harbin Institute of Technology, Shenzhen 518055, China}
\affiliation{Center for High Pressure Science and Technology Advanced Research, Shanghai 201203, China}

\author{Philip Dalladay-Simpson}
\affiliation{Center for High Pressure Science and Technology Advanced Research, Shanghai 201203, China}

\author{Ge Huang}
\affiliation{Center for High Pressure Science and Technology Advanced Research, Shanghai 201203, China}

\author{Zi-Yu Cao}
\affiliation{Center for Quantum Materials and Superconductivity (CQMS) and Department of Physics, Sungkyunkwan University, Suwon 16419, Republic of Korea}
\affiliation{Center for High Pressure Science and Technology Advanced Research, Shanghai 201203, China}

\author{Di Peng}
\affiliation{Center for High Pressure Science and Technology Advanced Research, Shanghai 201203, China}

\author{Federico Aiace Gorelli}
\affiliation{Center for High Pressure Science and Technology Advanced Research, Shanghai 201203, China}
\affiliation{National Institute of Optics (INO-CNR) and European Laboratory for Non-Linear Spectroscopy (LENS), Via N. Carrara 1, 50019 Sesto Fiorentino (Florence), Italy}

\author{Guo-Hua Zhong}
\affiliation{Shenzhen Institute of Advanced Technology, Chinese Academy of Sciences, Shenzhen 518055, China}
\affiliation{University of Chinese Academy of Sciences, Beijing 100049, China}

\author{Hai-Qing Lin}
\affiliation{Beijing Computational Science Research Center, Beijing 100193, China}

\author{Xiao-Jia Chen}
\email{xjchen2@gmail.com}
\affiliation{Center for High Pressure Science and Technology Advanced Research, Shanghai 201203, China}
\affiliation{School of Science, Harbin Institute of Technology, Shenzhen 518055, China}

\date{\today}

\begin{abstract}

Inspired by the high critical temperature in yttrium superhydride and the low stabilized pressure in superconducting cerium superhydride, we carry out four independent runs to synthesize yttrium-cerium alloy hydrides. The phases examined by the Raman scattering and x-ray diffraction measurements. The superconductivity is detected with the zero-resistance state at the critical temperature in the range of 97-140 K at pressures ranging from 114 GPa to 120$\pm$4 GPa. The maximum critical temperature of the synthesized hydrides is larger than those reported for cerium hydrides, while the corresponding stabilized pressure is much lower than those for superconducting yttrium hydrides. The structural analysis and theoretical calculations suggest that the phase of Y$_{0.5}$Ce$_{0.5}$H$_9$ has the space group $P6_3/mmc$ with the calculated critical temperature of 119 K, in fair agreement with the experiments. These results indicate that alloying superhydrides indeed can maintain relatively high critical temperature at modest pressures accessible by many laboratories.

\end{abstract}

\maketitle

Rare earth hydrides have been found to exhibit near room temperature superconductivity benefiting from the chemical pre-compression induced by the interaction between hydrogen and tetragen atoms \cite{Zurek,Pickard}. A typical example among them is the experimental discoveries of superconductivity at the critical temperature $T_c$ as high as 250-260 K at pressure of $\sim$170 GPa \cite{Somayazulu,Drozdov2} for clathrate LaH$_{10}$ with $Fm\bar{3}m$ structure \cite{Geball,Drozdov2} based on the early theoretical predictions \cite{fpeng,liuhy}. In parallel, yttrium hydrides as another attractive rare earth hydride system have drawn a lot of attentions for the predictions \cite{fpeng,liuhy,ywli} and experimental realization \cite{yshao,kongpp,Troyan,Snider1} of superconductivity. The experiments reported the $I4/mmm$-YH$_4$ phase with the maximum $T_c$ of 88 K at 155 GPa \cite{yshao}, the $Im\bar{3}m$-YH$_6$ phase with a similar $T_c$ near 220 K in the pressure range of 166-180 GPa \cite{kongpp,Troyan}, and the $P63/mmc$-YH$_9$ phase with the maximum $T_c$ $\sim$243 K at 201 GPa \cite{kongpp} or with the significantly high maximum $T_c$ $\sim$262 K at 182 GPa \cite{Snider1}. It is clear that the experimentally obtained YH$_x$ superconductors are only stable at extreme pressures (above $\sim$155 GPa). Meanwhile, theoretical \cite{Salke,lixin,fpeng,libin} and experimental \cite{Salke,lixin} efforts established that cerium superhydrides can be stabilized at low pressures below 100 GPa but also can be superconductive. The enhanced chemical pre-compression in CeH$_9$ was suggested to be associated with the delocalized nature of Ce 4$f$ electrons \cite{Jeon}. Recently, cerium superhydrides were found to exhibit superconductivity with the maximum $T_c$ of 115 K for $Fm\bar{3}m$-CeH$_{10}$ at 95 GPa and $T_c$ of 57 K for $P6_3/mmc$-CeH$_9$ at 88 GPa \cite{whchen}. Unlike La-H and Y-H systems, the stabilized pressure is dramatically reduced in the Ce-H compounds.

The exploration of high-$T_c$ superconductivity in superhydrides at low pressures or even ambient pressure is highly demanded. Compared with binary hydrides, ternary alloy hydrides possess diverse chemical compositions and thus provide more abundant structures for the operation with the advantages of different elements \cite{Cataldo,yfge,zzhang,xliang}. Interestingly, a series of lanthanum-yttrium ternary hydrides was reported to possess superconductivity with the maximum $T_c$ of 253 K at pressures of 170-196 GPa \cite{Semenok2}, indicating that hydrides can be stabilized in solid solutions at relatively low pressures. The superconductivity in lanthanum-cerium ternary superhydrides with $T_c$ $\sim$ 176 K was discovered at the moderate pressures \cite{hchen,jkbi}. Theoretical calculations demonstrate that a series of ternary superhydrides can hold high $T_c^{\prime}$s at relatively low stabilized pressure \cite{ysun,Cataldo,zzhang}. Therefore, exploring superconductivity in ternary hydrides towards high $T_c$ at modest pressures from the experimental side is an interesting direction. For such a purpose, we choose Y$_{0.5}$Ce$_{0.5}$ alloy and synthesize such alloy hydrides with the combined advantages of the low synthesized pressure in Ce-H and high $T_{c}$ in Y-H compounds.

Samples were prepared using a series of symmetric diamond-anvil cells (DACs). For the resistance measurements, the rhenium gaskets insulated by the mixture of cubic boron nitride (c-BN) and epoxy  were used to contain the sample at megabar pressures while isolating the platinum electrical leads. In detail, we employed the ammonia borane (NH$_3$BH$_3$ or AB) as the hydrogen source \cite{Somayazulu,whchen}. An Y$_{0.5}$Ce$_{0.5}$ slice with the thickness of about 1-2 $\mu$m was loaded next to AB in the gasket hole with the diameter of 40 $\mu$m and the thickness of 10-12 $\mu$m. 

\begin{figure}[tbp]
\includegraphics[width=\columnwidth]{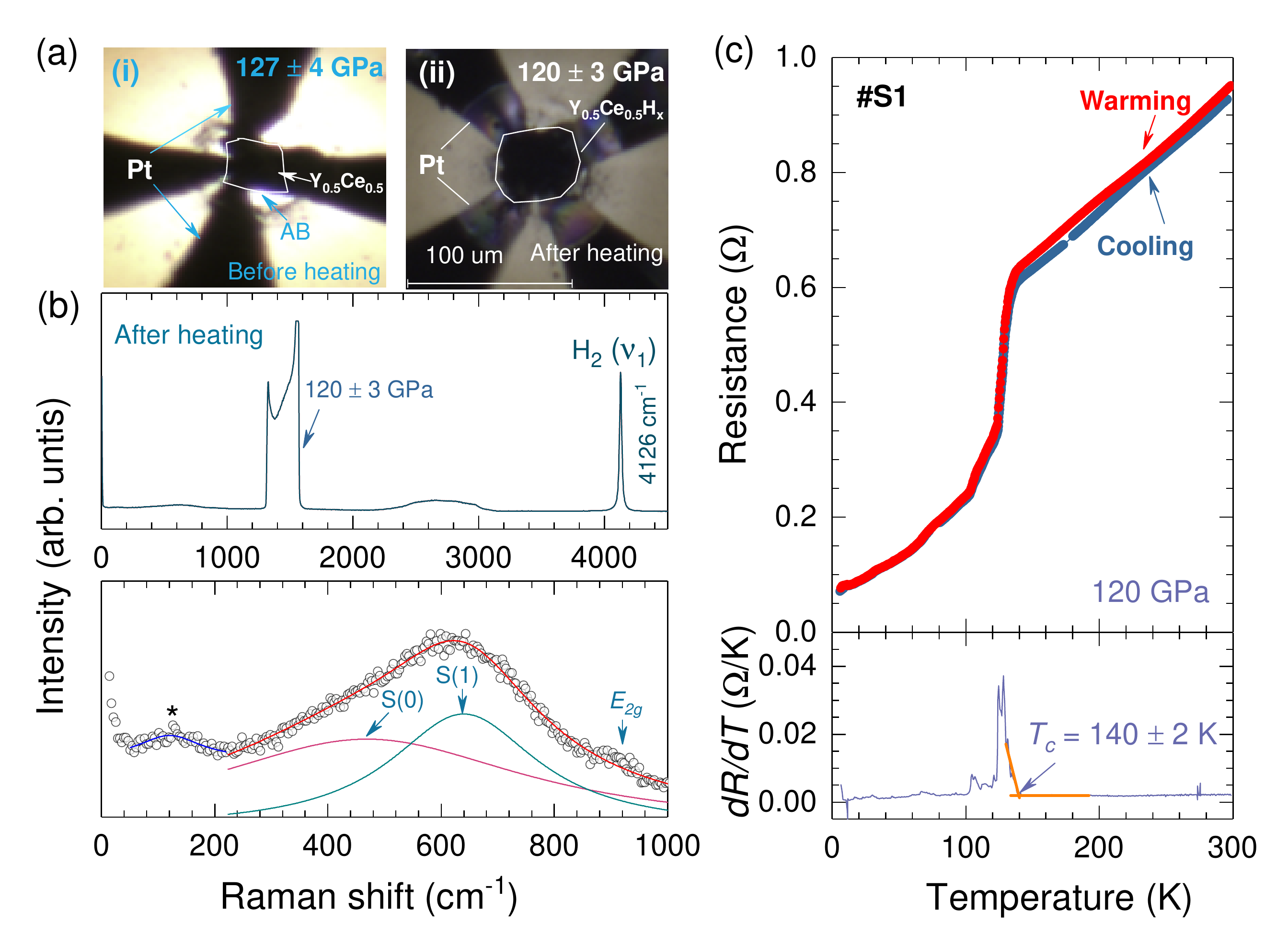}
\caption{Superconductivity in Y$_{0.5}$Ce$_{0.5}$ superhydrides in run 1. (a) Microphotographs of sample in DAC. (b) The Raman bands of the sample, diamond and H$_2$ after laser heating. (c) Temperature-dependent resistance.}
\end{figure}

The structural characterization was carried out at Shanghai Synchrotron Radiation Facility with a wavelength 0.6199 {\AA} and the size of the x-ray focus beam is less than 2 $\mu$m. The data was analyzed by using the software of Jana based on the Le Bail method \cite{jana,lebial}. The Raman scattering measurements were performed with the exciting laser (488 nm) and the beam size of the laser of about 5 $\mu$m. In all the experiments mentioned above, pressure was determined using the Raman shift of the stressed edge of the diamond peak \cite{bbaer}.

The electronic structures were calculated by using Vienna ab initio simulation package (VASP) \cite{Kresse1,Kresse2} with the density functional theory (DFT) \cite{Hohenberg,Kohn}. The generalized gradient approximation (GGA) of Perdew-Burke-Ernzerhof (PBE) functional was used for the exchange-correlation functional \cite{Perdew}. The cut-off energy of the plane-wave basis was set to 400 eV. The phonon dispersion, the electron phonon spectral function $\alpha^2F(\omega)$, electron-phonon coupling (EPC) parameter $\lambda$, and $T_c$ were carried out using the QUANTUM ESPRESSO (QE) package \cite{Giannozz} by employing the plane wave pseudopotential method and PBE exchange-correlation functional \cite{Perdew} together with the modified McMillan equation \cite{allen75}: $T_c$ = $\frac{\omega_{log}}{1.2}\exp[-\frac{1.04(1+\lambda)}{\lambda-\mu^\ast(1+0.62\lambda)}]$, where $\omega_{log}$ = $\exp[\frac{2}{\lambda}\int\ln(\omega)\frac{\alpha^2F(\omega)}{\omega}d\omega]$, $\lambda$ = 2$\int\frac{\alpha^2F(\omega)}{\omega}d\omega$, and $\mu^\ast$ being the Coulomb pseudopotential. The selected $k$-point mesh is 12 $\times$12$\times$12, and the $q$-point mesh is 3$\times$3$\times$3.  

\begin{figure}[tbp]
\includegraphics[width=\columnwidth]{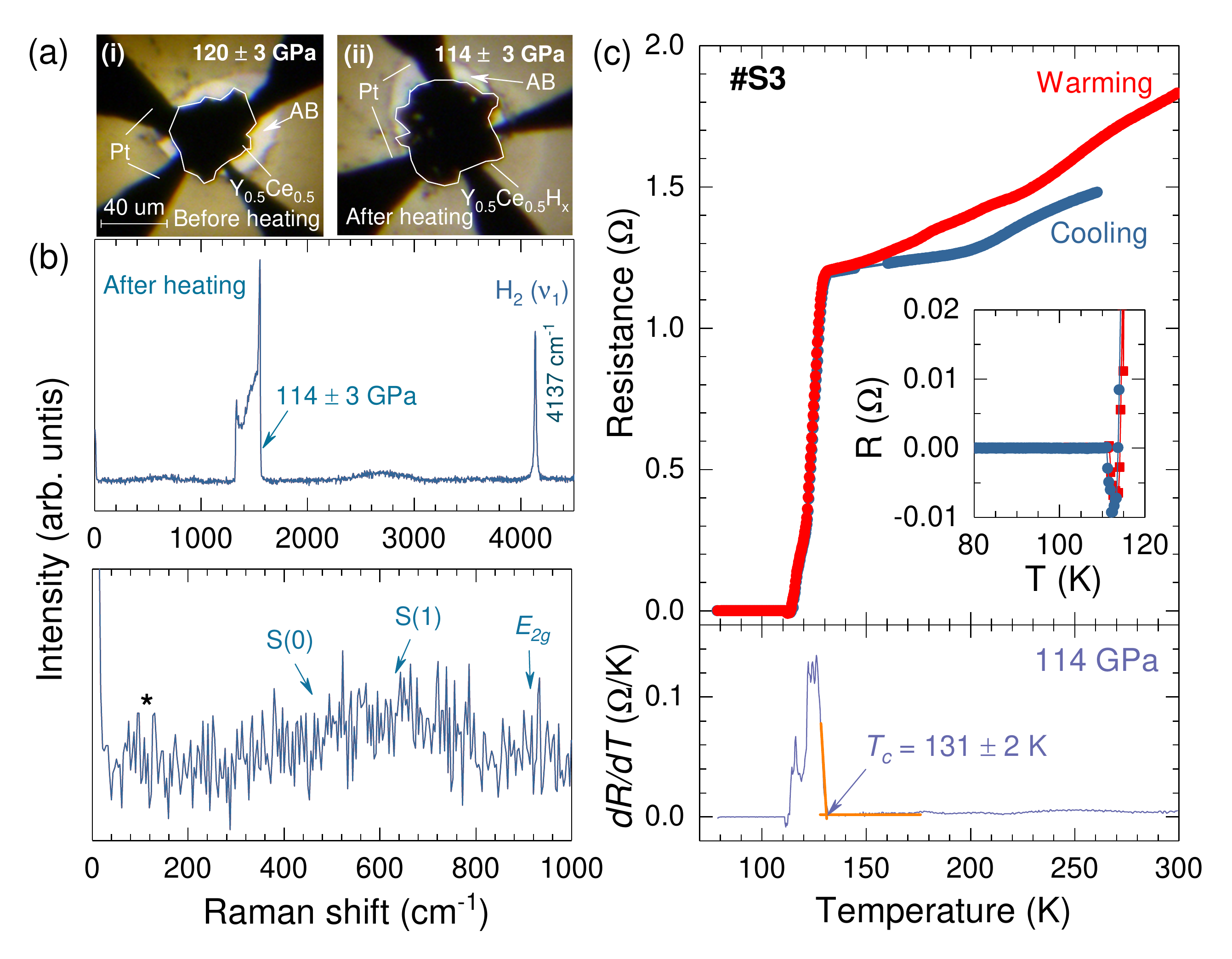}
\caption{Superconductivity in Y$_{0.5}$Ce$_{0.5}$ superhydrides in run 3. (a) Microphotographs for the synthetic progress. (b) The Raman bands of the sample, diamond and H$_2$ after laser heating. (c) Temperature dependence of the resistance.}
\end{figure}

For the first run, the pressure in DAC was increased to $\sim$127 GPa in one day after loading Y$_{0.5}$Ce$_{0.5}$ alloy and AB in the sample chamber [Fig. 1(a) (i)]. After the laser heating, the sample was visually expanded and became less reflecting [Fig. 1(a) (ii)]. Meanwhile, the pressure was found to drop to $\sim$120 GPa. The Raman vibron ($\nu_1$) of H$_2$ is detected with large intensity [Fig. 1(b)], indicating the presence of hydrogen in favor of the formation of Y$_{0.5}$Ce$_{0.5}$ superhydrides. The bands at about 469 and 640 cm$^{-1}$ are the rotational bands of H$_2$, named S(0) and S(1) respectively. The band around 918 cm$^{-1}$ is the optical phonon correlated with the transverse optical $E_{2g}$ mode of H$_2$ \cite{Zacs,Hemley}. The Raman band marked by an asterisk at about 120 cm$^{-1}$ should be the phonon mode of synthesized sample. The temperature dependence of the resistance ($R$) is presented in Fig. 1(c). A sharp $R$ drop can be seen around 140 K with a metallic behavior at higher temperatures. However, we have not reached the zero-resistance due to the incomplete reaction. 

To repeat the experiment, we performed other two experimental runs (\#S2 and \#S3). The pressure for the third run was increased to 120$\pm$3 GPa within two days. After laser heating, the sample expanded obviously, meanwhile the pressure dropped to 114$\pm$3 GPa. The lower panel of Fig. 2(b) shows the Raman bands at low frequencies, indicating the synthesis of Y$_{0.5}$Ce$_{0.5}$ superhydrides. In Fig. 2(c), the temperature-resistance ($T$-$R$) curve drops sharply below 130 K, indicating that the superconducting transition takes place. The zero-resistance state [inset of Fig. 2(c)] is the signature of superconductivity. The temperature dependent $dR/dT$ in the lower panel of Fig. 2(c) defines $T_c$ of 131$\pm$2 K.

\begin{figure}[tbp]
\includegraphics[width=\columnwidth]{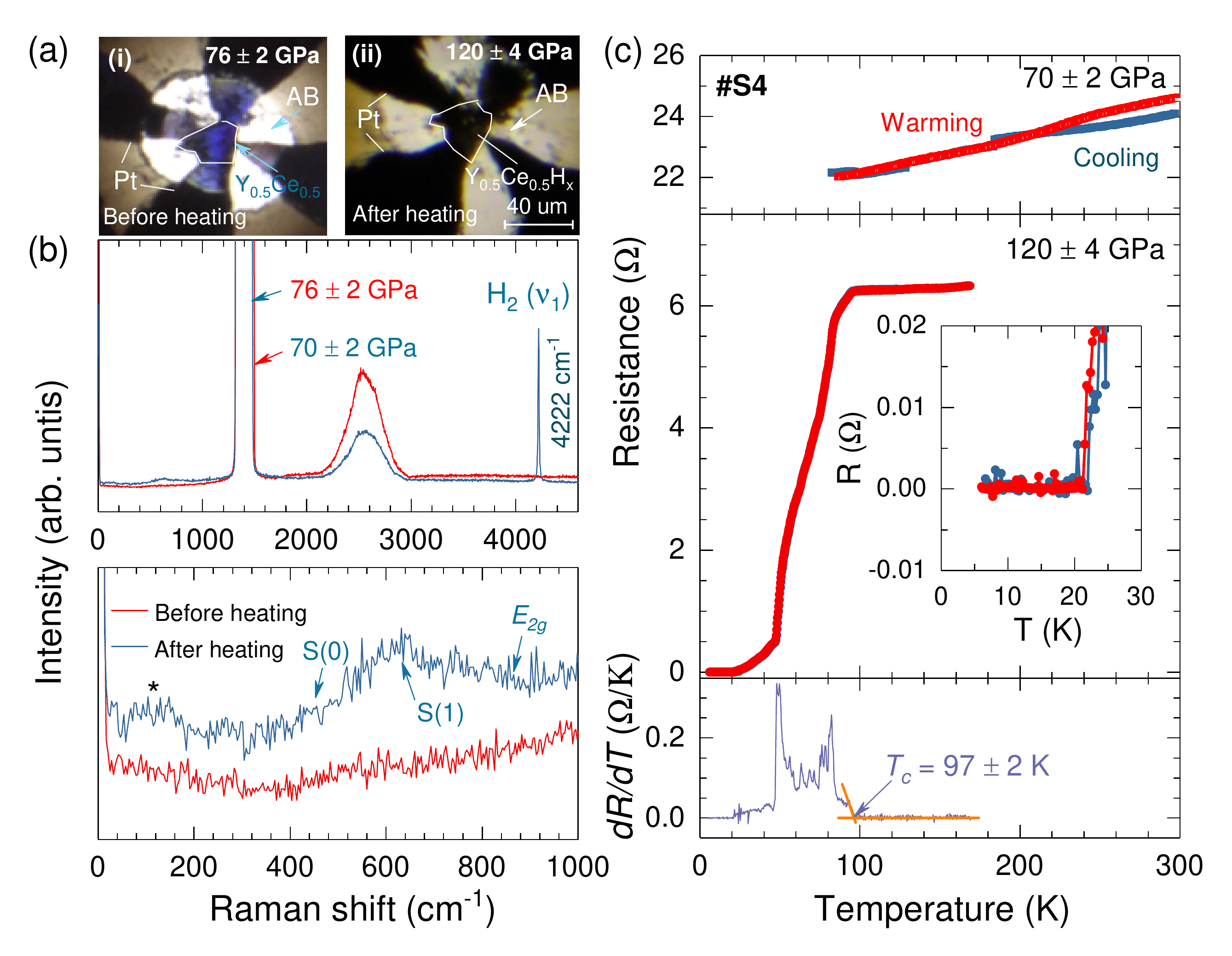}
\caption{Superconductivity in Y$_{0.5}$Ce$_{0.5}$ superhydrides in run 4. (a) Microphotographs for sample in DAC. (b) The Raman bands of the sample, diamond and H$_2$ before and after laser heating. (c) Temperature-dependent electrical resistance.}
\end{figure}

In the forth run, we first increased the pressure to 76$\pm$2 GPa [Fig. 3(a) (i)]. Obvious features for the synthesis of Y-Ce superhydrides appear such as the bands for H$_2$ and the optical band for the sample at low frequency [Fig. 3(b)]. However, only three Pt leads were connected with the sample after laser heating at the pressure of 70$\pm$2 GPa. Therefore, we only can detect the metallic behavior from the three-probe geometry [Fig. 3(c), upper panel]. In this circle, the superconducting transition was not detected probably due to the need of the lower temperature. When the pressure was increased to 120$\pm$4 GPa, all the Pt leads were found to contact with the sample. The middle panel of Fig. 3(c) presents the measured temperature-resistance curves at 120$\pm$4 GPa with the four-probe method. Unlike the observed behaviors for sample \#S1 and \#S3, a broad resistance drop can be clearly seen below 97 K for sample \#S4. High-$T_c$ superconductivity is confirmed by the zero-resistance state at low temperatures [inset of Fig. 3(c)]. Furthermore, the resistance drop of sample \#S4 has a wide temperature range from 97 to 23 K, indicating the possible non-homogeneous superconducting phase.

Synchrotron XRD results on sample \#S1 were used to determine the phase behavior [Fig. 4(a)]. Three phases with the space group of $P6_3/mmc$, $I4/mmm$ and $C2/m$ are analyzed. The refined lattice parameters at $\sim$120 GPa are $a$=3.64$\pm$0.01 {\AA}, $c$=5.41$\pm$0.02 {\AA}, $V$=31.07$\pm$0.26 {\AA}$^3$ for the $P6_3/mmc$ phase, $a$=2.81$\pm$0.02 {\AA}, $c$=5.86$\pm$0.02 {\AA}, $V$=23.18$\pm$0.41 {\AA}$^3$ for the $I4/mmm$ phase, and $a$=5.03$\pm$0.02 {\AA}, $b$=5.65$\pm$0.02 {\AA}, $c$=4.04$\pm$0.01 {\AA}, $V$=28.67$\pm$0.29 {\AA}$^3$ for the $C2/m$ phase, respectively. We compare the pressure dependent volume per unit of our data points and the reported results of binary Y-H and Ce-H in Fig. 4(b). For the same structure with equal hydrogen content, all the $P-V$ data points of YH$_x$ locate below CeH$_x$ due to the shorter H-H distance and smaller cationic radius in Y-H system \cite{whchen,kongpp,lixin}. The refined volume per unit of $P6_3/mmc$ phase for Y-Ce superhydrides gives the value between CeH$_9$ and YH$_9$, but closer to CeH$_9$. This indicates that the hydrogen content in the phase of $P6_3/mmc$ is 9. For the $I4/mmm$ phase in Y-Ce superhydrides, the volume per unit has larger value than that of YH$_4$, but is much close to that of CeH$_4$. Therefore, the hydrogen content in the phase of $I4/mmm$ is 4. However, the precise hydrogen content for the phase of $C2/m$ can not be determined due to the absence of volume for CeH$_7$. Knowing YH$_7$, the hydrogen content $x$ in the $C2/m$ phase might be 7. Similar results have been discussed in LaH$_x$ and LaH$_{6-7}$ with the same crystal structure \cite{Geball,hchen}. The crystal structures of Y$_{0.5}$Ce$_{0.5}$H$_{9}$ and Y$_{0.5}$Ce$_{0.5}$H$_{4}$ are shown in Fig. 4(c).

\begin{figure}[tbp]
\includegraphics[width=\columnwidth]{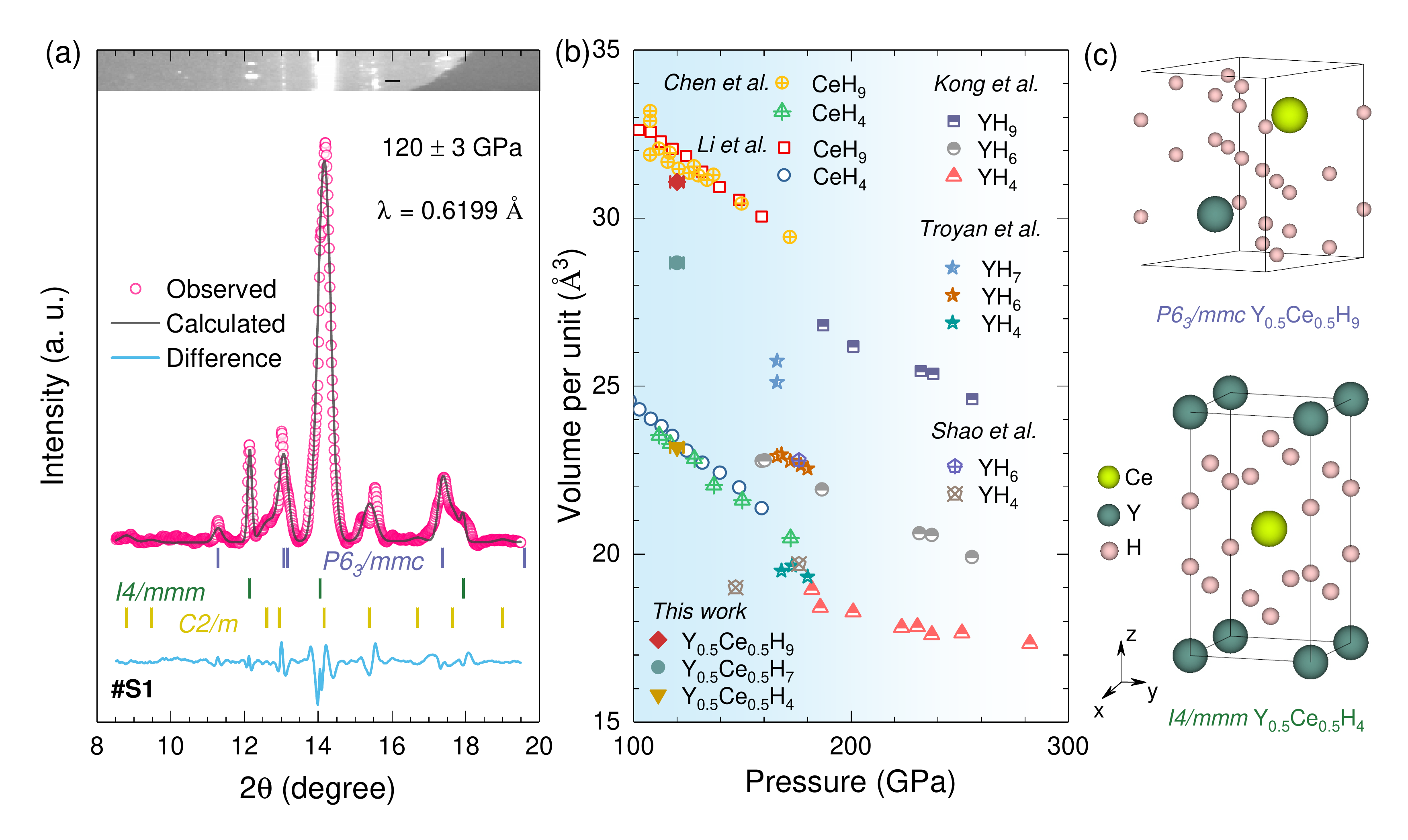}
\caption{Synchrotron x-ray diffraction of sample \#S1 at pressure of 120$\pm$3 GPa. (a) Le Bail refinements for the experimental data. (b) Volume per unit as a function of pressure. The experimental data are shown as solid symbols. The open and filled symbols are taken from Refs. \cite{yshao,kongpp,Troyan,lixin,whchen}. (c) The crystal structures of Y$_{0.5}$Ce$_{0.5}$H$_{9}$ and Y$_{0.5}$Ce$_{0.5}$H$_4$.}
\end{figure}

In order to understand the superconductivity with high $T_c$ in $P6_3/mmc$-Y$_{0.5}$Ce$_{0.5}$H$_9$, we calculated the electronic, phononic, and superconducting properties with the density functional theory (Fig. 5). The minimum pressure that can theoretically stabilize the structure $P6_3/mmc$-Y$_{0.5}$Ce$_{0.5}$H$_9$ is 180 GPa. As shown in Fig. 5(a), the $P6_3/mmc$ structure is metallic and shows a large density of states (DOS) near the Fermi level $E_F$. The large DOS is in favor of high-$T_c$ superconductivity. From the partial DOS shown in Fig. 5(a), we can see that the large DOS near $E_F$ is dominantly from the contribution of the H-$s$, Ce-$f$, and Y-$d$ orbitals. These results demonstrate that the high-$T_{c}$ superconductivity is a result of the joint contributions from all these elements in the hydrides. Meanwhile, the delocalized character of Ce-$f$ electrons provides the chemical pre-compression in such alloy hydrides. The calculated phonon dispersion, phonon density of states (PHDOS), $\alpha^2F$($\omega$), and $\lambda$($\omega$) of $P6_3/mmc$-Y$_{0.5}$Ce$_{0.5}$H$_9$ are given in Fig. 5(b). The optical phonon modes at high frequencies (above $\sim$390 cm$^{-1}$) are derived from the H atoms. The acoustic phonon modes with frequencies lower than $\sim$ 180 cm$^{-1}$ and the low-frequency optical modes between 145 and 320 cm$^{-1}$ are mainly attributed to the contributions of Y and Ce atoms. As seen from the Raman bands for synthesized Y-Ce superhydrides (Figs.  1-3), the phonon band at around 120 cm$^{-1}$ was assigned to the optical mode, which is the lowest optical phonon appeared at the experimentally observed regime from the calculations [Fig. 5(b)]. This mode is then taken as the signature of synthesized Y$_{0.5}$Ce$_{0.5}$H$_9$. 

\begin{figure}[tbp]
\includegraphics[width=\columnwidth]{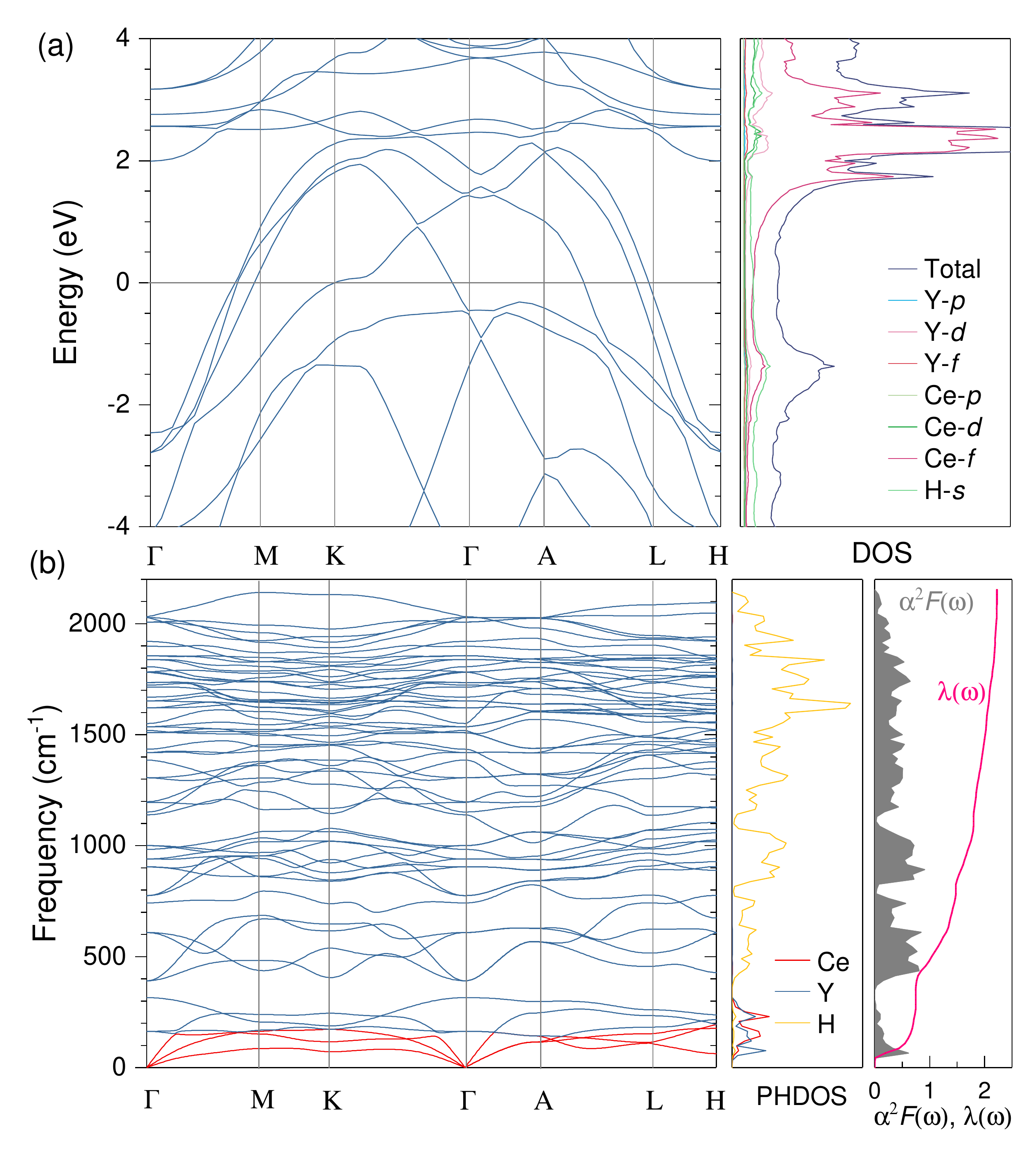}
\caption{The electronic (a) and phononic (b) properties of $P6_3/mmc$-Y$_{0.5}$Ce$_{0.5}$H$_9$ at 180 GPa. }
\end{figure}

From the phonon frequency dependence of $\alpha^2F$($\omega$) and integrated EPC constant $\lambda$($\omega$), we find that the contributions to $\alpha^2F$($\omega$) and $\lambda$($\omega$) arise from all three phonon modes including the Y-derived acoustic, the Ce-derived acoustic, and the H-derived optical modes. In addition, the wide high-frequency $\alpha^2F$($\omega$) is significantly higher than that on the narrow low-frequency side. Obviously, the high-energy H-derived vibrations dominate the total $\lambda$($\omega$) value. The calculations give a high EPC constant $\lambda$ of 2.23 at 180 GPa. By numerically solving the Eliashberg equation by choosing the typical values of 0.1-0.15 for $\mu^*$ \cite{Eliashberg}, we obtained $T_c$ of 104-119 K, in fair agreement with our experiments.

\begin{figure}[tbp]
\includegraphics[width=\columnwidth]{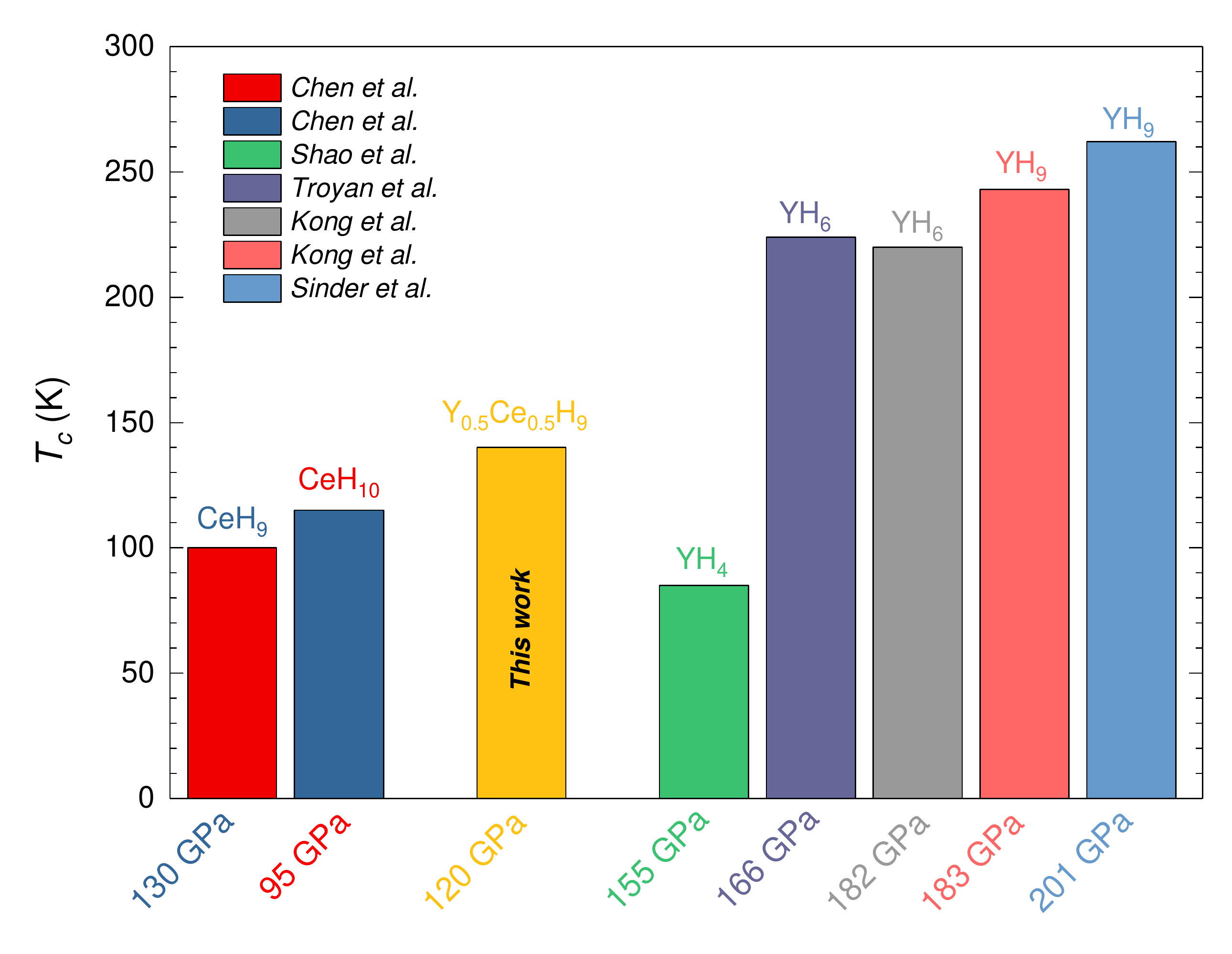}
\caption{Comparison of the maximum $T_c$ values in cerium, yttrium, and Y$_{0.5}$Ce$_{0.5}$ alloy hydrides at various pressures.}
\end{figure}

In the end, we briefly summarize the maximum $T_c$ values of cerium hydrides and yttrium hydrides reported in the literature with the comparison of ours (Fig. 6). Yttrium hydrides (YH$_6$ and YH$_9$) show the advantages of superconductivity with high $T_c$ values. However, the extremely high pressure makes them difficult to be operated \cite{yshao,kongpp,Troyan,Snider1}. In contrast, cerium hydrides (CeH$_{10}$ and CeH$_9$) largely reduce the stabilized pressure but with the expense of the $T_c$ value \cite{Salke,lixin,whchen}. It should be noticed that our synthesized Y$_{0.5}$Ce$_{0.5}$H$_9$ possesses lower stabilized pressure than those for yttrium hydrides and higher $T_c$ than those of cerium hydrides. Thus, the successfully synthesized Y$_{0.5}$Ce$_{0.5}$H$_9$ points out the promising direction for exploring high $T_c$ at low stabilized pressure in alloy hydrides.

In summary, we have successfully synthesized Y$_{0.5}$Ce$_{0.5}$ hydrides by laser heating at high pressures in four runs (samples, \#S1-S4). The obtained temperature dependence of the resistance provides the evidence for superconductivity in Y$_{0.5}$Ce$_{0.5}$H$_{9}$ with the highest $T_c$ of 140$\pm$2 K at 120$\pm$3 GPa. The structural analysis revealed the coexistence of $P6_3/mmc$-Y$_{0.5}$Ce$_{0.5}$H$_9$ with $C2/m$-Y$_{0.5}$Ce$_{0.5}$H$_{7}$ and $I4/mmm$-Y$_{0.5}$Ce$_{0.5}$H$_4$. The theoretical calculations indicated that $P6_3/mmc$-Y$_{0.5}$Ce$_{0.5}$H$_9$ is dynamically stable at the studied pressure range and possesses comparable $T_{c}$ values with the experiments. These findings hope to advance the search of alloy hydrides with high-$T_c$ at relatively moderate pressure.

\begin{acknowledgments}
The work was supported by the Basic Research Program of Shenzhen (Grant No. JCYJ20200109112810241), the Shenzhen Science and Technology Program (Grant No. KQTD20200820113045081), the National Post-doctoral Program for Innovative Talents (Grant No. BX2021091), and the National Key R\&D Program of China (Grant No. 2018YFA0305900).
\end{acknowledgments}

\end{document}